\documentclass[10pt,letterpaper,twocolumn]{article}

\usepackage{ol}
\usepackage{hyperref}
\usepackage{amsmath,amssymb}

\begin{document}
\twocolumn[
\title{Measurement of the complex dielectric constant of \\a single gold nanoparticle}

\author{Patrick Stoller$^{\dagger}$, Volker Jacobsen, Vahid Sandoghdar$^*$}
\affiliation{Laboratory of Physical Chemistry, ETH Zurich, 8093
Zurich, Switzerland}

\begin{abstract}
A differential interference contrast microscopy technique, which
employs a photonic crystal fiber as a white-light source, is used
to measure both the real and imaginary parts of the complex
dielectric constant of single 10 and 15\,nm gold nanoparticles
over a wavelength range of 480 to 610\,nm. Noticeable deviations
from bulk gold measurements are observed at short wavelengths and
for individual particles even after taking into account
finite-size surface damping effects.
\end{abstract}


] Gold nanoparticles exhibit a plasmon resonance peak in the
optical absorption and scattering spectrum that is absent from the
reflectivity spectrum of bulk gold~\cite{kreib1995}. Bohren and
Huffman~\cite{bohren1983} present comparisons between absorption
and scattering cross-sections calculated from the measured bulk
dielectric constant using Mie theory~\cite{mie1908} and the
results of direct measurements on ensembles of nanoparticles
smaller than 20 nm. While they find good agreement in general,
they observe that the measured absorption peak is broader and
lower than that predicted by Mie theory. This discrepancy can be
partly explained by a particle-size dependence of the dielectric
constant~\cite{kreib1995}, the main correction being due to an
additional damping that arises when the conduction electron mean
free path of about 10\,nm becomes comparable to the particle size.
However, previous studies have noted that a systematic discrepancy
between directly measured and calculated absorption values
persists even after taking this correction into
account~\cite{dorem1964,granq1977,gu2005}. Since these
measurements have been done in ensembles of gold particles, it is
not possible to completely exclude the finite particle size
distribution as an explanation of observed discrepancies. As a
result, a few groups have developed methods for the spectroscopy
of single gold nanoparticles with diameters below
20\,nm~\cite{lindfo2004,Berciaud:05,Muskens:06}. None of these,
however, has been able to measure independently the real and
imaginary parts of the dielectric constant of a single
nanoparticle. In this work we combine the detection principle
presented in Ref.~\cite{lindfo2004} with a modified version of
differential interference contrast
microscopy~\cite{batch1989,taube1991,matsuo2001} to achieve this.

\begin{figure}[htb]
\centerline{\includegraphics[width=8cm]{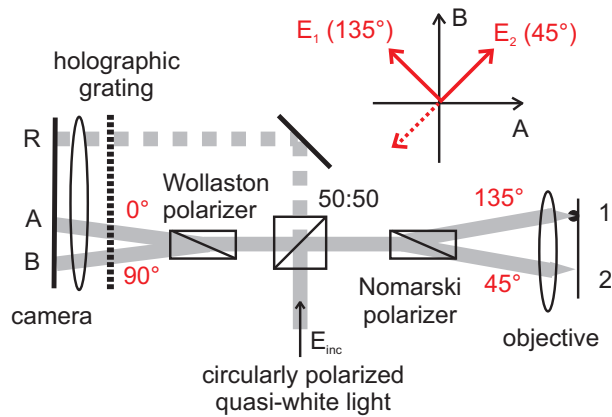}}
\caption{Illustration of the phase sensitive detection principle.
The polarization angles are specified next to each beam. See text
for details.} \label{setup}
\end{figure}

The interferometric detection scheme used is sketched in
Fig.~\ref{setup}. Quasi-white supercontinuum light generated in a
photonic crystal fiber~\cite{lindfo2004} was collimated and then
linearly polarized using a Glan-laser polarizer. A phase
compensator, consisting of a wavelength independent Fresnel-rhomb
polarization rotator followed by an achromatic quarter wave plate
(QWP), allowed adjustment of the phase between light polarized at
45$^{\circ}$ and 135$^{\circ}$ angles with respect to the fast
axis of the QWP. A Nomarski polarizer, with its fast and slow axes
oriented at 45$^{\circ}$ to those of the QWP was used to generate
two beams labelled 1 and 2, separated by an angle of
0.014$^{\circ}$. A plan-apochromatic oil-immersion objective
(Zeiss 1.4\,NA, 63x, DIC) focused the beams into two
non-overlapping spots with orthogonal polarizations on the upper
surface of a standard glass cover-slip. The light returning from
spots 1 and 2 was recombined by the Nomarski, passed through a
confocal pinhole, and subsequently split by a Wollaston polarizer
(splitting angle of 0.5$^{\circ}$) into two linearly polarized
components with polarization angles oriented at 0$^{\circ}$ and
90$^{\circ}$ with respect to the fast axis of the QWP. As
indicated in Fig.~\ref{setup}, if spot 1 contains a particle, the
optical field $E_{1}$ from this surface is projected onto the
positive B (90$^{\circ}$) axis and onto the negative A
(0$^{\circ}$) axis. The field $E_{2}$ from the spot without the
particle is projected onto the positive A and B axes. Thus, the
detectors A and B record the signals
$|E_{\textrm{1}}-E_{\textrm{2}}|^2$ and
$|E_{\textrm{1}}+E_{\textrm{2}}|^2$, respectively. The field
$E_{1}$ consists of the field reflected at the interface and the
field $E_{p}\propto\alpha E_{inc}$ scattered by the particle where
$\alpha$ is the polarizability of the particle.

Let us now define a complex quantity $X(\omega)$ as
\begin{eqnarray}
X(\omega)\equiv\frac{2}{\pi\varepsilon_{m}d^3}\alpha(\omega)=
\frac{\varepsilon_{p}(\omega)-\varepsilon_{m}}
{\varepsilon_{p}(\omega)+2\varepsilon_{m}} \label{def}
\end{eqnarray}
where $\omega$ is the frequency of light and $\varepsilon_{p}$,
$\varepsilon_{m}$ are the dielectric constants of the gold
nanoparticle and the surrounding medium ($\varepsilon_{m}=2.30$),
respectively. $X$ represents the part of the complex
polarizability that is not directly dependent on the particle
size. Following Ref.\cite{matsuo2001}, it can be shown that the
real and imaginary components of $X$ are given by the following
relations:
\begin{align}
        X_{re}&=\frac{\lambda}{2Cd^3}\:\left(A-B\right)
    \label{XRE}\\
        X_{im}&=\frac{\lambda}{Cd^3}\:\left(1 -
        \sqrt{A+B-1-\tfrac{1}{4}(A-B)^2}\right)
\label{XIM}
\end{align}where $d$
is the particle diameter, $\lambda$ is the wavelength. $C$ is a
real proportionality constant that is independent of particle size
and wavelength but includes an apparatus function. We calibrated
$C$ by measuring the signal from 50\,nm silica beads
(Polysciences) with a real index of refraction of
$n=1.46$\cite{hecht1979}. So starting with the detector signals
\emph{A} and \emph{B}, we can calculate $X_{re}$ and $X_{im}$, and
then solve for the dielectric constant of the gold particles:
\begin{align}
\varepsilon_{re}&=\frac{1-2X_{im}^2+X_{re}-2X_{re}^2}{X_{im}^2+(1-X_{im})^2}\;\varepsilon_{m}
\label{epsre}\\
\varepsilon_{im}&=\frac{3X_{im}}{X_{im}^2+(1-X_{re})^2}\;\varepsilon_{m}.
\label{epsim}
\end{align}

The gold nanoparticles (British Biocell) were spin-coated onto
cover slips at a density of less than about 1 particle/$\mu$m$^2$.
We conducted careful studies to ensure that our sample preparation
resulted in single particles, and not
aggregates~\cite{lindfo2004}. A drop of immersion oil was placed
on top of the cover-slip to match its index of refraction. This
provides an optically uniform surrounding medium and eliminates
any systematic effects on the plasmon spectrum due to the
substrate. The two beams from the Wollaston traversed a
holographic diffraction grating (Kaiser, HFG\,550) and were imaged
onto a Peltier-cooled CCD camera using an achromatic lens. A
spectral resolution of about $\sim$\,1\,nm was readily achieved.
As indicated in Fig.~\ref{setup}, a reference beam labelled $R$
from the beam splitter cube was also sent through the diffraction
grating and imaged onto the same camera. When the sample was
scanned a camera image with the three spectra $A$, $B$, and $R$
was acquired at each pixel. Spectra $A$ and $B$ were divided by
$R$ at each pixel to correct for intensity and spectral
fluctuations in the incident light. Particles passing through
either of the two focal spots on the sample led to a change in
spectra $A$ and $B$. We note that the spectra $A$, $B$ and $R$
showed small deviations even without the presence of a particle
due to slight chromatic imperfections of the optical components
used. To correct for this, the on-particle spectra were normalized
further using a background spectrum obtained at a nearby point on
the sample without any particles.

\begin{figure}[htb]
\centerline{\includegraphics[width=8.5cm]{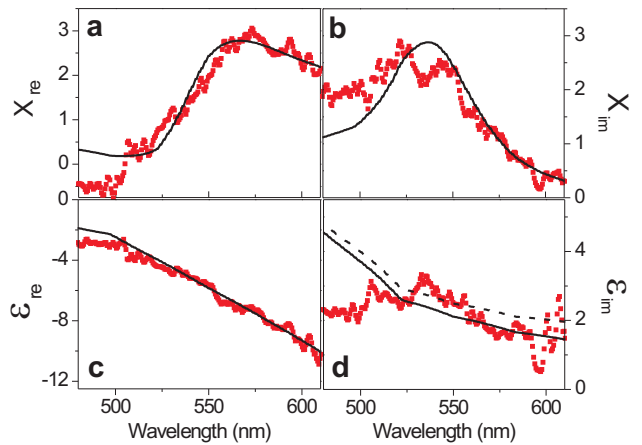}}
\caption{Plot of X$_{re}$ (a) and X$_{im}$ (b) as a function of
wavelength for a single 15\,nm gold nanoparticle. The solid line
shows the calculated value for a spherical particle in the
quasi-static approximation using the bulk dielectric constant of
gold. Plot of the real (c) and the imaginary (d) part of the
dielectric constant calculated from (a) and (b). The solid curve
shows the bulk dielectric constant and the dashed curve shows this
quantity corrected for the size-limited mean free conduction
electron path.} \label{15nm_single}
\end{figure}

In Figs.~\ref{15nm_single}a) and b), we have plotted $X_{re}$ and
$X_{im}$ as measured on a single 15\,nm particle. The solid curves
show the calculated fits from Eqs.~(\ref{XRE}) and (\ref{XIM})
using the bulk dielectric constant of gold taken from the
measurements~\cite{johnson1972,innes1987}, leaving the particle
diameter as the only fit parameter. We obtained 10$\pm$1\,nm for
the nominally 10\,nm particles and 12$\pm$1\,nm for the nominally
15\,nm particles, while the manufacturer specifies a standard
deviation in the particle diameter of $\sim$\,10\% based on
transmission electron microscope studies. The symbols in
Figs.~\ref{15nm_single}c) and d) illustrate $\varepsilon_{re}$ and
$\varepsilon_{im}$ calculated from the data in the upper panels
using Eqs.~(\ref{epsre}) and (\ref{epsim}), respectively whereas
the solid curves represent the literature values of bulk
gold\cite{johnson1972}. We find a clear disagreement between these
plots. We note that Muskens et al.~\cite{Muskens:06} have recently
obtained satisfactory fits for the absorption plasmon spectra of
small single gold nanoparticles by using $\varepsilon_{m}=2.04$
and by taking into account the surface damping~\cite{bohren1983}.
Indeed we would also obtain a better fit for this value of
$\varepsilon_{m}$. However, we have chosen to fix the refractive
index of the surrounding medium to $\varepsilon_{m}=2.3$ because
we believe this is a well known parameter of the system.

\begin{figure}[htb]
\centerline{\includegraphics[width=8.5cm]{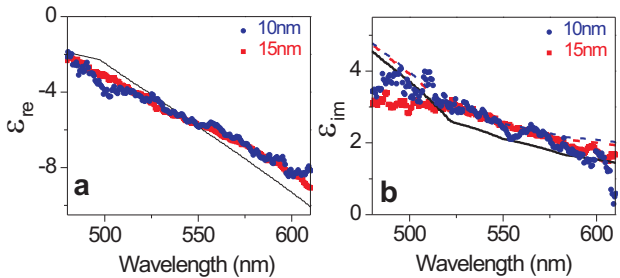}}
\caption{Plot of (a) the real and (b) the imaginary part of the
dielectric constant calculated by averaging over fifteen 10\,nm
particles (blue) and thirteen 15\,nm particles (red). The solid
curve shows the bulk dielectric constant and the dashed curves
show the same quantity corrected for the size-limited mean free
conduction electron path in 10 and 15\,nm particles.}
\label{10nm_and_15nm_average}
\end{figure}

The particles we use are likely not perfectly spherical. By using
a model in Eq.~(\ref{XRE}) and Eq.~(\ref{XIM}) that takes into
account spheroidal particles, it is possible to obtain somewhat
better agreement between measured and calculated values of
$X_{re}$ and $X_{im}$ than that shown in Figs.~\ref{15nm_single}a)
and b). The fit results obtained using such calculations give
average aspect ratios of 0.74 for both the 10 and 15\,nm
particles. We have analyzed transmission electron microscope
images of many 10\,nm gold nanoparticles and have obtained an
estimate of the aspect ratio between 0.8 and 1.0 for all of the
particles measured. We thus rule out particle ellipticity as the
sole cause for the deviation observed in Fig.~\ref{15nm_single}a)
and b). In order to compensate for shape dependent effects on the
measured dielectric constant, we averaged over multiple particles
of each nominal size (this also reduces the fluctuations seen in
individual spectra). Fig.~\ref{10nm_and_15nm_average} illustrates
the average of $\varepsilon_{re}$ and $\varepsilon_{im}$ from
thirteen 10\,nm particles (blue) and fifteen 15\,nm particles
(red). For wavelengths longer than about 600\,nm, where the
plasmon resonance is weak, $X_{re}$ and $X_{im}$ become noisy and
cause large fluctuations in $\varepsilon_{re}$ and
$\varepsilon_{im}$ (see Eqs.~(\ref{epsre}, \ref{epsim})). In
Figs.~\ref{10nm_and_15nm_average}a) and b), the solid black curves
show the real and imaginary values of the bulk dielectric
constant. The blue and red dashed curves in
Fig.~\ref{10nm_and_15nm_average}b) display this quantity after
taking into account the limited mean free path of the conduction
electrons in 10\,nm and 15\,nm particles, respectively. Here we
added the surface damping correction term
$\tfrac{3}{4}(\omega_p^2\lambda^3v_f)/(4\pi^3c^3d)$ to
$\varepsilon_{im}$ of bulk gold, where $\omega_p$ is the plasma
frequency of gold, $c$ the speed of light in vacuum, and $v_f$ the
electron velocity at the Fermi surface\cite{bohren1983}. The data
for 10\,nm particles is noisier because the signal from these
particles is about 3 times smaller than that of 15\,nm
particles~\cite{lindfo2004}.

Despite the residual intensity and spectral noise in the
white-light continuum, it is clear from
Fig.~\ref{10nm_and_15nm_average} that there is good agreement
between the real part of the dielectric constant measured in bulk
and that measured on 10 and 15\,nm particles. Reasonably good
correspondence exists between bulk and small particle values of
the imaginary part of the dielectric constant in the range of the
plasmon resonance, i.e. from about 510 to 580\,nm. This agreement
is even better when one takes into account the mean free path
limitation. However, there is marked disagreement between the bulk
and small particle values of $\varepsilon_{im}$ for wavelengths
shorter than about 510\,nm. This discrepancy is related to the
facts that the measured spectra of $X_{im}$ show a broader and
shallower peak than what is calculated and that the measured
spectrum of $X_{re}$ dips below the calculated spectrum (see
Fig.~\ref{15nm_single}a,b). The broadening of $X_{im}$, which
corresponds to a broadening in the absorption spectrum, has
already been observed in previous studies on ensembles of gold
particles\cite{dorem1964}. Imperfections of and impurities in the
gold nanoparticles and chemical interface damping\cite{hendr2003}
can both result in additional broadening, but this is difficult to
verify for chemically synthesized particles that are coated with
surfactants, as used in this work.

To our knowledge, we have reported on the first measurements of
both the real and the imaginary parts of the complex dielectric
constant of single gold nanoparticles over a broad spectral range.
Intensity and spectral noise in the white-light continuum limited
the measurement to particles larger than about 10 nm. We are
currently working to improve these so we can perform quantitative
measurements on single clusters provided by well-controlled
fabrication methods\cite{kreib2001}. Further measurements would
shed light on the deviations of the dielectric constant of small
nanoparticles from that of bulk matter.

This work was performed within the Innovations-Initiative (INIT)
program of ETH Zurich on Composite Doped Metamaterials.\\
$^*$~vahid.sandoghdar@ethz.ch,~www.nano-optics.ethz.ch.
\\ $^{\dagger}$ Present address:\ Institute of Applied Physics,
University of Bern, Switzerland.

\end{document}